\documentstyle[sprocl,epsfig]{article}

\def\be{\begin{equation}}
\def\ee{\end{equation}}
\def\bea{\begin{eqnarray}}
\def\eea{\end{eqnarray}}
\begin{document}

\title{THE FORMATION OF THE SEYFERT X-RAY CONTINUUM}

\author{P. O. PETRUCCI}

\address{Laboratoire d'Astrophysique de Grenoble, 414 rue de la
Piscine,\\ BP 38041 Grenoble, France\\E-mail: petrucci@obs.ujf-grenoble.fr} 


\maketitle\abstracts{}

%====================================================================%
\section{Introduction and generalities}
%====================================================================%

%====================================================================%
\subsection{The UV-X-$\gamma$ spectrum of Seyfert galaxies}
%====================================================================%
The typical broad band UV-X-$\gamma$ ray spectrum of a Seyfert galaxy,
here NGC 5548 has been reported in Fig. \ref{specsey} This spectrum has
been obtained from quasi simultaneous observations of different
satellites indicated on the Figure \cite{mag98}. We clearly
see different components:
\begin{figure*}[b!]
\vskip -0.7 true cm
\begin{center}
\centerline{\epsfig{figure=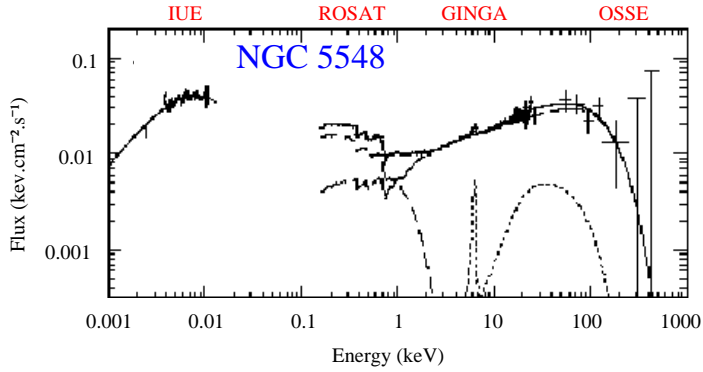,
width=0.8\textwidth}}
\end{center}
\vskip -0.3 true cm
\caption{Typical broadband UV-X-$\gamma$ ray spectrum of a Seyfert galaxy.}
\label{specsey}
\end{figure*}
\begin{itemize}
\item A strong emission in the UV, the UV bump, generally interpreted as
the thermal emission of an accretion disc.
\item A strong emission in the X-ray band which can be generally fitted
by a cut-off power law shape with a power law photon index $\Gamma\simeq
1.9$ \cite{nan94} and a cut-off energy $E_c$ near 200-300 keV
\cite{mat99,per02}
\item A neutral fluorescence Fe$_{K\alpha}$ line near 6.4 keV \cite{nan97}
\item A bump peaking near 30 keV \cite{pou90}
\item A soft component below 2 keV, in excess to the extrapolation of the
high energy power law, called the soft excess \cite{tur89}
\end{itemize}

\begin{figure*}[h!]
\vskip -0.7 true cm
\begin{center}
\centerline{\epsfig{figure=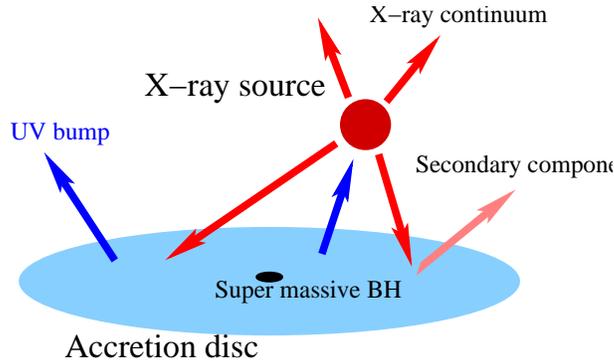, width=8cm}}
\end{center}
\vskip -0.3 true cm
\caption{Schematic view of the reprocessing/upscattering model}
\label{repmodel}
\end{figure*}

These different components are generally interpreted in the
reprocessing/upscattering model framework. The scheme of this model is
reported in Fig. \ref{repmodel}. It supposes the presence of 2 phases, a
cold one, the accretion disc, which produces the UV emission, and a hot
one, the corona, a plasma of energetic particles, supposed to radiate
X-rays by Compton upscattering the UV photons coming from the
disc. Inversely, part of the X-ray emission of the corona is absorbed by
the cold material and reprocessed in X-ray. Finally, part of the X-ray
emission ($\sim$ 10\%) is Compton reflected on the disc surface and give
birth to the fluorescent Iron line and the reflection bump. The origin of
the soft excess is still unclear, it may be the hard tail of the disc
emission.

The nature of the high energy source(s), the geometry of the disc-corona
configuration and the origin of the variability, which is important in
Seyfert galaxies, are still unknown. The presence of a high energy
cut-off in most of the Seyfert (Matt 2000) and the lack of any
annihilation line have favored a thermal nature for the corona even if it
does not rule out the non-thermal hypothesis \cite{pet01}. Recent results
on spectral variability (cf. section \ref{sectnat}) appear to be more
conclusive in agreement with the thermal interpretation. We will thus
focus below on the thermal solution.

%====================================================================%
\subsection{The thermal Comptonization spectrum}\label{subsec:comp}
%====================================================================%
As said previously, the X-ray emission in Seyfert is believed to be
produce by Compton process of UV seed photons on a thermal population of
electrons. For a given geometry, the thermal Comptonization spectral
shape depends on few parameters i.e. the temperature and optical depth of
the corona, $kT_e$ and $\tau$, and the temperature of the soft photon
field $kT_{bb}$.

\begin{figure*}[h]
\begin{center}
\centerline{\epsfig{figure=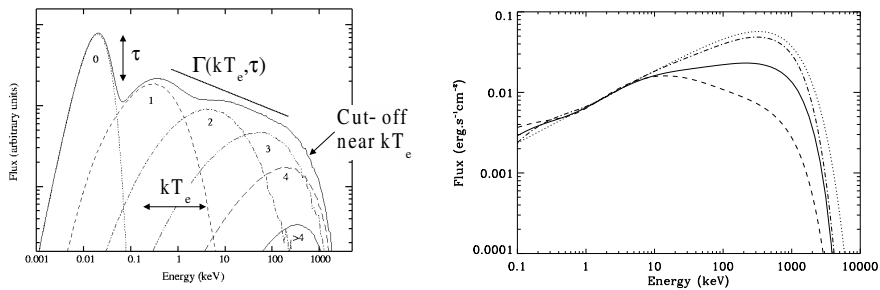,
width=1\textwidth}}
\end{center}
%\begin{center}
%\begin{tabular}{cc}
%\hspace*{-3.5cm}\epsfig{figure=petruccifig3.eps, width=0.5\textwidth}
%&\hspace*{-6cm}\epsfig{figure=petruccifig4.eps,width=0.5\textwidth}
%\end{tabular}
%\end{center}
\caption{Left: A thermal Comptonization spectrum in isotropic
geometry. Right: comparison of different Comptonization spectra for
different geometry: slab (solid line), hemisphere (dashed line), sphere
(i.e. isotropic geometry, dot-dashed line).  The values of $\tau$ and $kT_e$ in each
case have been chosen so that the spectra have roughly the same slope at
low energy. We clearly see the anisotropy break present for anisotropic
geometries. In dotted line we have plotted a simple cut-off power law
approximation.}
\label{thercompspec}
\end{figure*}

It generally has a cut-off power law shape (cf. Fig. \ref{thercompspec}a)
and it is often approximate by a spectral function of the type
$F(E)\propto E^{-\Gamma}\exp (-E/E_c)$. This may be however a very crude
approximation especially in the case of anisotropic geometries. Indeed,
the Compton process is relatively sensible to the anisotropy of the seed
photon field and the X-ray emission may also be strongly anisotropic. In
the case of a slab geometry for example, the X-ray photons of the first
Compton scattering order are preferentially emitted backward toward the
disk than toward the observer. It thus produces a break (the anisotropy
break) in the observed X-ray spectrum as shown in
Fig. \ref{thercompspec}b. In this case, the use of a cut-off power-law
shape may lead to erroneous parameter values and to wrong physical
interpretations \cite{pet00,pet01}.

It is interesting to note that in the case of radiative equilibrium
between the UV (i.e. the cooling) and the X-ray (i.e. the heating)
sources, the temperature and optical depth of the corona follow an
univocal relationship, which depends however on the geometry of the
disc-corona configuration \cite{haa93,ste95}. This relationship roughly
corresponds to a constant Compton parameter $y = \displaystyle \left[
4\frac{kT_e}{mc^2}+16\left(
\frac{kT_e}{mc^2}\right)^2\right](\tau+\tau^2)$.

%====================================================================%
\section{Geometry of the corona}
%====================================================================%
Geometries more ``photon-starved'' than the simple slab corona are
clearly required by the data \cite{haa94,dim99,pet01,bia00,der02}.

\begin{figure*}[h!]
\vskip -0.7 true cm
\begin{center}
\centerline{\epsfig{figure=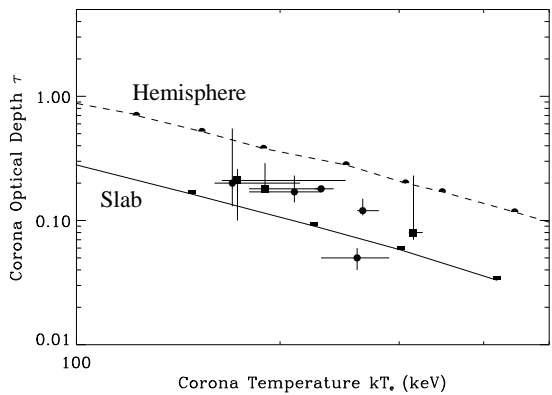,width=8cm}}
\end{center}
\vskip -0.3 true cm
\caption{Best fitting results of BeppoSAX observation of a sample of
Seyfert galaxies using a realistic Comptonization model. The theoretical
expectation for a slab and hemispherical corona geometry have been over
plotted in solid and dashed line respectively. From Petrucci et
al. (2001).}
\label{photonstarvedgeom}
\end{figure*}

Fitting BeppoSAX observations of a sample of Seyfert galaxies with a
realistic Comptonization model in slab geometry, we were able to
constrain the temperature and optical depth of the corona. The
corresponding values have been reported in Fig. \ref{photonstarvedgeom}
together with the theoretical relationship expected in the case of a slab
(solid line) and hemispherical (dashed line) geometry in radiative
equilibrium (cf. Sect. \ref{subsec:comp}). The data points tend to fall
preferentially in between the two line, indicating that the real
configuration is more complex than these two ideal case.

%====================================================================%
\section{High energy variability}
%====================================================================%
\label{sectnat}
%====================================================================%
\subsection{Long time scale variability}
\begin{figure*}[h]
\begin{center}
\centerline{\epsfig{figure=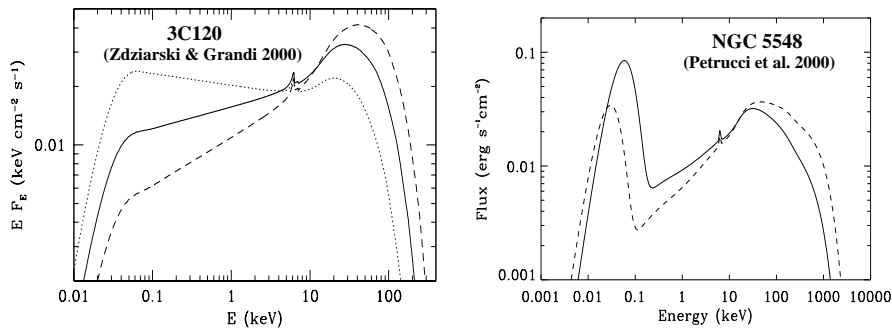,
width=1\textwidth}}
\end{center}
%\begin{center}
%\begin{tabular}{cc}
%\hspace*{-3.5cm}\epsfig{figure=petruccifig6.eps,width=0.5\textwidth}
%&\hspace*{-6cm}\epsfig{figure=petruccifig7.eps,width=0.5\textwidth} 
%\end{tabular}
%\end{center}
\caption{Best fit models of BeppoSAX data. Left: 3C120. Right: NGC
5548. The spectra are clearly pivoting at high ($\simeq$ 10 keV) energy.}
\label{pivotspec}
\end{figure*}

%====================================================================%
There is growing observational evidence, at least on long (i.e. $> 1
day$) time scale, that the X-ray spectrum of Seyfert softens (i.e. the
photon index $\Gamma$ increases) as the 2-10 keV flux increases
\cite{per86,yaq93,lee99,pet00,vau01,zdz01,pag02,der02}. We also generally
observed larger amplitude variability in the soft X-ray band than in the
hard X-ray band \cite{mar01}. These different results indicate
that the continuum is mainly pivoting at high energy as indeed observed
in some cases (cf. Fig. \ref{pivotspec}).

This support the idea that the X-ray spectral variability (at least on
long time scale) is most likely due to a change of the cooling i.e. the
X-ray spectral variability is driven by the UV seed photons, as expected
in the upscattering model interpretation.

\begin{figure*}[h]
\begin{center}
\centerline{\epsfig{figure=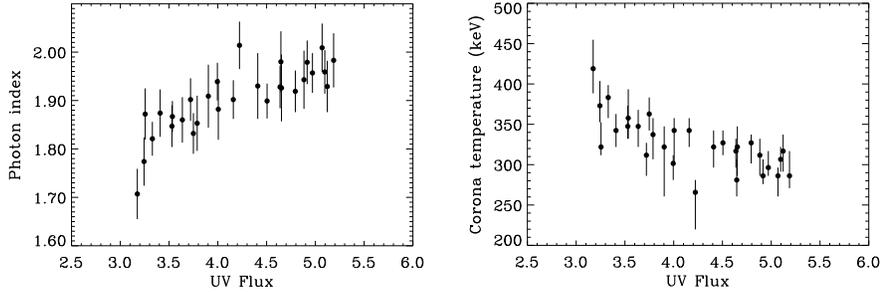,
width=1\textwidth}}
\end{center}
%\begin{center}
%\begin{tabular}{cc}
%\hspace*{-3.5cm}\epsfig{figure=petruccifig8.eps, width=0.5\textwidth}
%&\hspace*{-6cm}\epsfig{figure=petruccifig9.eps,width=0.5\textwidth}
%\end{tabular}
%\end{center}
\caption{Correlation between the X-ray photon index and the UV flux
(left) and anti-correlation between the corona temperature and the UV
flux during the 30 days simultaneous IUE/XTE observation of NGC
7469. These results strongly support thermal Comptonization models.}
\label{correl}
\end{figure*}

Recent results from the analysis of the IUE/XTE observation of NGC 7469
strongly agree with this conclusion. NGC 7469 has indeed been observed
simultaneously by these 2 satellites during 30 days. Nandra et
al. (2000), fitting the XTE data with a simple power law for the
continuum, found a clear correlation between $\Gamma$ and the UV flux
(cf. Fig. \ref{correl}). A re-analyze of these data with a realistic
Comptonization model\cite{pet02} shows that the increase of the UV flux
goes with a decrease of the corona temperature (cf. Fig. \ref{correl}),
supporting a thermal nature for the corona.

%====================================================================%
\subsection{Short time scale variability}
%====================================================================%
The rapid variations of the X-ray flux (on time scale of a few hours or
less) are generally not present in UV (cf e.g. Nandra et al. 1998 in the
case of NGC 7469). This clearly indicates that there is an X-ray
variability process distinct from the one occuring on longer time
scale. If on the latter case, the variability precursor appears to be the
UV source, on smaller time scale it has certainly something to do with
the process that heats the corona, for instance reconnection instability
in a magnetically dominated corona \cite{dim98,mer01}.

Nandra \& Papadakis (2001) have recently shown that the power density
spectrum (hereafter PSD) of NGC 7469 hardens for higher energy, i.e. the
slope of the PSD computed in the 2-4 keV band was larger than the one in
the 10-15 keV band. Moreover these differences between the PSD are in the
sense that the high frequency ($> 10^{-4}$ Hz) power spectrum shows more
power at high energies (the opposite is true however at low frequency in
agreement with the upscattering model expectations when the seed photon
source is variable). This result also indicates that there is a mechanism
unrelated to the seed photons that dominates variability on small time
scale. As already proposed above, the corona may consist of magnetic
flares. The hardening of the PSD could then be explained by the presence
of hotter flares in the inner regions of the disc since they would mainly
account for the high-energy photons and be also more rapidly variable due
to their smaller size scale.

%====================================================================%
\section{Variability and correlation with spectral properties between
objects}
%====================================================================%
In general, more luminous objects are less variable in X-ray
\cite{gre93,nan97,mar01} and this effect is even stronger for smaller
time scale\cite{mar01} . It indicates that the variability
characteristics between objects differ mainly by their time scale rather
than by their amplitude.  A natural way to explain this behavior would be
that more luminous object have larger black hole mass that is larger
length scale and thus less rapid variability.

However, some type of Seyfert galaxies, the Narrow Line Seyfert, show,
for a given luminosity, a larger variability than a standard Seyfert
galaxy \cite{tur99,lei99}. According to our previous remark, the larger
variability of these objects suggests that they possess a central black
hole with a relatively small mass. Consequently, they have to possess a
larger accretion rate to explain their luminosity.

This thus suggest that Seyferts follow a $\dot{M}/{M_{BH}}=\dot{m}$
classification in the sense that the Broad line Seyferts may have small
$\dot{m}$ and the Narrow line Seyferts may have large
$\dot{m}$\cite{pou95} . Many of the details are unclear however. For
instance, the more variable objects have in general a softer spectra
\cite{bol96,tur99} which clearly show that the Comptonizing corona have
different physical characteristics between objects.

%====================================================================%
\section{Corona formation}
%====================================================================%
Different models have tried to explain the formation of the corona and to
reproduce the observations and the different correlation described
above. The corona may then be produced by disk evaporation and/or
particles may be accelerated trough magnetic reconnection like in the
solar corona \cite{mey00,mer02,liu02}. Such models manage to reproduce
the main characteristics of the high energy emission discussed above like
the correlation between $\Gamma$ and the X-ray flux, the temperature
gradient in the inner regions of the corona, and the dependence of the
spectral properties with $\dot{m}$ from the existence or not (depending
on $\dot{m}$) of a radiation pressure dominated area in the central
region of the accretion disk. For more details the reader can refer to
the references given above.

%====================================================================%
\section*{Acknowledgments}
%====================================================================%
I would like to thanks the organizing comitee for having invited me to
present this talk and for the great interest of this meeting.


\section*{References}
\begin{thebibliography}{99}

\bibitem{bia01} Bianchi, S. et al. 2001, A\&A, 376, 77 

\bibitem{bol96} Boller, T., Brandt, W.~N., \& Fink, H.\ 1996, A\&A, 305,
53

\bibitem{der02} De Rosa, A. et al. A\&A, 387, 838

\bibitem{dim98} Di Matteo, T.\ 1998, MNRAS, 299, L15

\bibitem{geo01} Georgantopoulos, I.~\& Papadakis, I.~E.\ 2001, MNRAS,
322, 218

\bibitem{gre93} Green, A.~R., McHardy, I.~M., \& Lehto, H.~J.\ 1993,
MNRAS, 265, 664

\bibitem{haa93} Haardt, F.  \& Maraschi, L.  1993, ApJ, 413, 507 

\bibitem{haa94} Haardt, F. 1994, PhD dissertation, SISSA, Trieste (H94)

\bibitem{lam00} Lamer, G., Uttley, P., \& McHardy, I.~M.\ 2000, MNRAS,
319, 949

\bibitem{lee99} Lee, J.~C. et al. 1999, MNRAS, 310, 973

\bibitem{lei99} Leighly, K.~M.\ 1999, ApJS, 125, 297

\bibitem{liu02} Liu, B.~F. et al.\ 2002, ApJ, 575, 117

\bibitem{mag98} Magdziarz, P. et al. 1998, MNRAS, 301, 179

\bibitem{mar01} Markowitz, A.~\& Edelson, R.\ 2001, ApJ, 547, 684

\bibitem{mat99} Matt, G., Proceeding of the conference ``X-ray
Astronomy '999. Stellar Endpoints, AGN and the Diffuse Background'',
September 6-10, Bologna, Italy. To appear in Astrophysical Letters and
Communications

\bibitem{mey00} Meyer, F., Liu, B.~F., \& Meyer-Hofmeister, E.\ 2000,
A\&A, 361, 175

\bibitem{mer01} Merloni, A.~\& Fabian, A.~C.\ 2001, MNRAS, 321, 549

\bibitem{mer02} Merloni, A.~\& Fabian, A.~C.\ 2002, MNRAS, 332, 165

\bibitem{nan94} Nandra, K.~\& Pounds, K.~A.\ 1994, MNRAS, 268, 405

\bibitem{nan97} Nandra, K. et al. 1997, ApJ, 477, 602

\bibitem{nan01} Nandra, K.~\& Papadakis, I.~E.\ 2001, ApJ, 554, 710

\bibitem{nan98} Nandra, K. et al.\ 1998, ApJ, 505, 594

\bibitem{nan00} Nandra, K. et al.\ 2000, ApJ, 544, 734 

\bibitem{pag02} Page, K.~L. et al.\ 2002, MNRAS, 330, L1

\bibitem{per86} Perola, G.\ C.\ et al. 1986, ApJ, 306, 508

\bibitem{per02} Perola, G.~C. et al. 2002, A\&A, 389, 802 

\bibitem{pet00} Petrucci, P.\ O.\ et al. 2000, ApJ, 540, 131

\bibitem{pet01} Petrucci, P.~O., Henri, G., \& Pelletier, G.\ 2001, A\&A,
374, 719

\bibitem{pet02} Petrucci, P.\ O.\ et al. 2002, A\&A,

\bibitem{pou90} Pounds, K.\ A. et al. 1990, Nature, 344, 132


\bibitem{pou95} Pounds, K.~A., Done, C., \& Osborne, J.~P.\ 1995, MNRAS,
277, L5

\bibitem{ste95} Stern, B. E. et al. 1995, ApJL, 449, L13

\bibitem{tur89} Turner, T.~J.~\& Pounds, K.~A.\ 1989, MNRAS, 240, 833

\bibitem{tur99} Turner, T.~J. et al. 1999, ApJ, 524, 667

\bibitem{vau01} Vaughan, S.~\& Edelson, R.\ 2001, ApJ, 548, 694

\bibitem{yaq93} Yaqoob, T. et al. 1993, MNRAS, 262, 435

\bibitem{zdz01} Zdziarski, A.~A.~\& Grandi, P.\ 2001, ApJ, 551, 186



\end{thebibliography}
\end{document}